\documentclass[10pt,a4paper]{article}
\usepackage{t1enc}
\usepackage{ae} 
\usepackage{aecompl} 
\usepackage[latin1]{inputenc}
\usepackage[english]{babel}
\usepackage{natbib,amssymb,amsmath,amscd,amstext}
\usepackage{amsfonts}
\usepackage{amsthm}
\usepackage{graphicx}
\usepackage{psfrag}
\usepackage{epsfig}
\usepackage{stmaryrd}
\usepackage{epstopdf}
\newcommand{\dsz}{\ensuremath{d_{\text{SZ}}}}
\newcommand{\dhz}{\ensuremath{d_{\text{HZ}}}}
\newcommand{\lt}{\ensuremath{l_{\text{T}}}}

\newcommand{\taum}{\ensuremath{\tau_{-\epsilon}}}
\newcommand{\taup}{\ensuremath{\tau_{\epsilon}}}
\numberwithin{equation}{section}

\begin{document}
\title{The lipid bilayer at the mesoscale: a physical continuum model} \author{P.L. Wilson\footnote{Department of Mathematics and Statistics, University of Canterbury, Private Bag 4800, Christchurch 8140, New Zealand. Corresponding author. Email: p.wilson@math.canterbury.ac.nz}, H. Huang\footnote{Department of Mathematics and Statistics, York University, Toronto, Canada M3J 1P3. Email: hhuang@mathstat.yorku.ca}, S. Takagi\footnote{Department of Mechanical Engineering, The University of Tokyo, Tokyo, Japan. Email: takagi@mech.t.u-tokyo.ac.jp}} \maketitle

\begin{quote}
We study a continuum model of the lipid bilayer based on minimizing the free energy of a mixture of water and lipid molecules. This paper extends previous work by \citet{bap} in the following ways. (a) It formulates a more physical model of the hydrophobic effect to facilitate connections with microscale simulations. (b) It clarifies the meaning of the model parameters. (c) It outlines a method for determining parameter values so that physically-realistic bilayer density profiles can be obtained, for example for use in macroscale simulations. Points (a)-(c) suggest that the model has potential to robustly connect some micro- and macroscale levels of multiscale blood flow simulations. The mathematical modelling in point (a) is based upon a consideration of the underlying physics of inter-molecular forces. The governing equations thus obtained are minimized by gradient flows via a novel numerical approach; this enables point (b). The numerical results are shown to behave physically in terms of the effect of background concentration, in contrast to the earlier model which is shown here to not display the expected behaviour. A ``short-tail'' approximation of the lipid molecules also gives an analytical tool which yields critical values of some parameters under certain conditions. Point (c) involves the first quantitative comparison of the numerical data with physical experimental results.
\end{quote}

\section{Introduction}\label{sec:intro}
We study the potential utility of a continuum paradigm \citep{bap}
of the lipid bilayer as a mesoscale filter in multiscale simulations
of blood flow. The two main points are first that we introduce a
more physical model of the hydrophobic effect into the paradigm (for
consistency with microscale simulations), and second that we
investigate the role and meaning of the paradigm's parameters (to
connect with macroscale simulations) by performing the first quantitative comparison of numerical solutions of the paradigm with physical experimental data, and in so doing provide a method for determining parameter values.

As is well-known, the cell is the fundamental element of all living matter. The activity of the cell sustains life and the cell itself is sustained by a metabolism which utilizes the mass transfer through its membrane. The cell membrane is composed of lipid molecules with proteins and other components floating in it. The dynamics of this lipid bilayer membrane become especially important in the case of dispersed components in the blood, such as red blood cells, white blood cells, platelets and so on, because the deformation dynamics of these membranes directly affect the mass transfer in the blood. These membranes are often modelled as hyperelastic due to the presence of a cytoskeleton. On the other hand, a liposome --- which is composed of lipid bilayers only --- is modelled as a two-dimensional fluid membrane, because the membrane lipid molecules can easily move laterally within the bilayer. Liposomes are used as drug delivery agents and artificial oxygen-carriers in blood. Nevertheless, all these cases are defined by the lipid bilayer and the behavior of the bilayer is responsible for the mass transfer through the cell membranes. Hence, the modelling of the lipid bilayer membrane from a molecular level through to the continuum level is highly anticipated to predict the mass transfer behavior in blood \citep{mchedlishvilimaeda2001,boryczkoetal2003,sugiietal2005,vaidyaetal2007}.

Blood flow is a complex and major aspect of the growing field of
multiscale simulations of the human body
\citep{aytonetal2002, sugiietal2007, boal2002, mouritsen2005}. It is multiphase, with red blood cells
(RBCs), white blood cells, platelets, artificial drug delivery
agents (DDAs) and other components dispersed in a water-like plasma. It involves multiphysics, with blood components having
viscoelastic properties and mutually interacting while also
exchanging mass and undergoing chemical reactions. Moreover, it is
multiscale; quantum mechanics at the smallest length and time scales
upwards through other physics at intermediate scales, and continuum
mechanics at the longest scales are all required to properly
understand, predict, and control the behaviour and properties of
blood. Advances in computer processing speed and storage capacity have made multiscale simulations possible, highlighting the need to understand the computational and physical interactions between the scales.

Here we focus on one crucial intermediate scale, or mesoscale, at
which both molecular physics and continuum mechanics are important.
This is at the membrane which surrounds RBCs and some DDAs, being
typically only two molecules in thickness but extending laterally
for several micrometres. Understanding how membrane composition affects
deformability, and how deformation affects the mass-transfer
properties of RBCs and DDAs, are key to the multiscale modelling of
blood flow, since the blood vessels in which occur the greatest mass
transfer from these components are also those in which they undergo
the greatest deformations due to large shear forces and mechanical
constriction in narrow capillaries. One significant challenge for
multiscale simulations almost independent of large increases in
computational power and storage, is the need to reliably transfer
information from one level of the simulation to another. Presently
\citep{young2001}, this is mostly done \emph{ad hoc} and offline, and may
omit crucial mesoscale dynamics.

The membrane is a bilayer of lipid molecules (lipids) \citep{mouritsen2005}. The dipolar
charge distribution in the lipid ``head'' groups make them able to
bond with water molecules, while ``tails'' are indifferent to the
presence of water. This amphiphilic nature leads to the hydrophobic
effect which physically forms the bilayer and gives it integrity \citep{chandler2005}. In
this paper, we consider bilayers composed of one type of lipid.

\citet{bap} base their continuum paradigm, herein called the ``BP paradigm'', on
the mesoscopic dynamics framework of \citet{fraaije1993}, minimizing
a free energy for a system of lipid and water molecules. Formally,
the intrinsic free energy of the system is minimized with respect to
a constraint that the (unobservable) distribution of the molecules
generates the (observable) continuous volume fractions, thus
assuming that the microstate has relaxed to equilibrium over the
relatively long time scale of the continuous description.

We introduce a new model of the inter-molecular interactions into
the BP paradigm which in contrast to the original model closely
represents the underlying physics of the hydrophobic effect. A
further improvement is a term $\beta$ to control the decay rate of
the interaction strengths. Our new model and other improvements
anticipate being able to determine the values of more of the model
parameters from microscale simulations (or first principles). Since
tests not reported here show that both models have very similar
computational costs, we concentrate largely on our new model. Our point is not that our model has computational advantages or performs better under certain conditions, but rather that it has a more physical basis and so enables a better connection to microscale simulations or first principles. It is therefore encouraging that our more physical model gives results qualitatively similar to the original model of \cite{bap}. Moreover, we show here that a key characteristic of the bilayers produced by the new model depends on a model parameter in a physically realistic way, in contrast to the original model (\S\ref{subsub:cmc}).

Understanding the paradigm's parameters and the numerical roles they
play is important for connecting to macroscale simulations and
physical experiments. By considering the analytical origins of the
paradigm, we show that some of those parameters which are at first sight
physical (as opposed to purely numerical) are seen to be largely
numerical, in that they cannot be directly connected \emph{a priori}
with physical measurements. Properties of numerical bilayers must be
compared \emph{a posteriori} with physical properties in order to
set some parameter values. The numerical bilayers are obtained by a method of solution new to the paradigm, and are for the first time quantitatively compared with physical bilayers.

In more detail, the system of water molecules and heads and tails of
lipids has a free energy split into an ideal part roughly
corresponding to the Helmholtz free energy, involving only
connectivity interactions, and a non-ideal part representing
inter-molecular interactions.
Lipid structure and configuration are therefore explicitly
represented. \citet{bap} formulate a non-ideal part of the free
energy with a term reflecting the (global) compressibility of the
system and another modelling the (local) hydrophobic interactions;
the modelling of the hydrophobic interaction term is an open
question and is not inherent to the formalism of
\citet{fraaije1993}. In this paper we introduce a new model of the
hydrophobic interactions which captures the physics underlying the
hydrophobic effect responsible for the formation and integrity of
ordered states of amphiphiles, based on the following discussion.

Liquid water is a dynamic hydrogen bond network in which each water
molecule forms up to four hydrogen bonds with its neighbours. The
non-zero dipole moment of lipid head groups makes them able to
accept hydrogen bonds from water molecules (but unable to donate a bond
to each other): they are hydrophilic. By contrast, the hydrophobic
lipid tail groups are unable to form hydrogen bonds, although
thermodynamic and electrostatic interactions between water molecules
and tail groups occur, but in liquid water at room temperature the
hydrogen-bond energy is typically an order of magnitude stronger
than such interactions \citep{immergut1991,mouritsen2005}.

A cavity with a structured ``surface'' in the hydrogen bond network
forms around a hydrophobic moeity, causing a decrease in the entropy
of the system \citep{kronbergetal1995}. An entropic force acts to
gather together hydrophobic moieties so as to minimize the disruption
to the hydrogen bond network. The physical origin of this
hydrophobic effect is that water molecules close to a sufficiently
large hydrophobic moiety no longer participate in four hydrogen
bonds; with no attractive force towards the hydrophobic moeity,
these molecules' remaining bonds now draw them away from the moiety.
It is thus because lipid head groups can be nodes in the hydrogen
bond network while tail groups cannot that bilayers and other
structured lipid assemblies form, and this is the basis of our model.

Of the physical parameters, we take the system to be incompressible,
leaving the effects of the compressibility parameter $p$ to future
work. The relative tendency of heads and waters to form hydrogen
bonds is modelled here by the new parameter $\gamma$ which is set as
unity in this paper: the effects of $\gamma$ are also left to future
work, while we simply note here that $\gamma$ moves the paradigm beyond
modelling heads as attached water molecules. Herein we investigate
the effects of the key physical parameters $\alpha, \epsilon, \beta,
c_0$, and $m$. Respectively these represent temperature effects,
lipid head-tail group separation distance, decay of the interaction
strength, and a ``background concentration'' and ``excess mass'' of
lipids (these last two terms, inherited from
\citet{bap}, are clarified in this paper).

The paper is structured as follows. We introduce into the BP
paradigm a new model of the inter-molecular interactions in
\S\ref{subsec:modmodel} which forms the basis of the numerical
solutions. The parameters and lipid model are discussed here.
Euler-Lagrange equations, whose solution minimizes the free energy
functional, are derived in \S\ref{subsec:el}, and a novel numerical
approach to solving them is given in \S\ref{subsec:numerics}, along
with a sample numerical result. All solutions and discussions are
based on a one-dimensional model. The ``smoothed'' nature of the
paradigm and the choice of lipid model are discussed in
\S\ref{subsub:lipmod}.

\S\ref{sec:phys} connects the BP paradigm to physical \emph{in
vitro} measurements of lipid bilayers. The summary of bilayer
properties in \S\ref{subsec:bilayer} is used in
\S\ref{subsec:newinterp} to describe how numerical solutions can be
calibrated to physical data. Numerical solutions are calibrated in
this way for the parameters of interest in \S\S\ref{subsub:cmc},2,3.
In particular, a short-range interaction (or ``short-tail'') approximation based on the
new parameter $\beta$ is introduced in \S\ref{subsub:epsibeta}, and
its effects studied analytically and numerically. This work enables
guidelines on the choice of parameter values to be given in in
\S\ref{subsec:params}. The conlusions are in \S\ref{sec:conc}.

\section{The new model: derivation and numerics}\label{sec:model}

\subsection{The new model}\label{subsec:modmodel}
The original model of the hydrophobic interaction acted to move
tails away from heads and waters by penalising proximity between
them, mimicking the \emph{effect} of the hydrophobic force but not
the underlying \emph{cause}, namely the uneven force distribution on
water molecules in close proximity to lipid tails described above.
Our approach, in direct contrast to the original model, is to
promote water-water and water-head (but not head-head) proximity,
modelling the hydrogen bond network, and effectively ignore the
hydrophobic tails in direct imitation of the underlying physics. The
relative strength of the water-water bonding preference to the
water-head bonding preference is controlled by a parameter $\gamma$,
meaning that heads are no longer the unphysical ``attached waters''
of \citet{bap}.

Formally, our system comprises ``waters'', each represented by a
single ``bead'', and ``lipids'', each
represented by a ``head'' bead and a ``tail''
bead\footnote{All beads are of zero dimension.} connected by a rigid massless rod of
length $\epsilon$. The one-dimensional model has two lipid groups
aligned in the $x$-direction, normal to the bilayer plane; one group
whose tails, having normalized density $u(x)$, point in the
positive $x$-direction, and the other with tails of normalized density
$v(x)$ pointing in the negative $x$-direction. The head beads of the
first group have normalized density $\taum u(x)=u(x+\epsilon)$, and
similarly for the second. Water beads have normalized density
$w(x)$. The lipids have here been chosen as the simplest allowed in
the paradigm; we later (\S\ref{subsub:lipmod}) argue that more
complex models mostly would not improve the utility of the model as
a mesoscale ``filter'' in a multiscale simulation.

The total free energy of the system consists of three parts in the
form
\begin{equation}\label{eq:fullfree}
\begin{split}
E=T\int [\eta(u)+\eta(v)+\eta(w)]+\frac{p}{2}\int(1-u-v-\taum
u-\taup v-w)^2\\ + \alpha\int w\hat{\kappa}*[w+\gamma(\taum u+\taup
v)]\ .
\end{split}
\end{equation}
The new model differs from the original in the third part; the
meanings of all three parts, the parameters and variables follow.

The first term, favouring spreading, represents the entropy of the
system, in which $\eta(s)=s\log s$ for non-negative $s$ and
$\eta(s)=\infty$ otherwise, and where $T$ is the temperature of the
system. The second term is a potential energy due to
compressibility, where $p$ is the system pressure.

The new form of the third term, modelling directly the underlying
physics of the hydrophobic effect, involves a water-water term
$w\hat{\kappa}*w$ and a water-head term $w\hat{\kappa}*\gamma(\taum
u+\taup v)$, where $*$ indicates convolution in the form
\begin{equation}\label{eq:conv}
(f\hat{\kappa}*g)(x)=\int f(x)\hat{\kappa}(x-y)g(y)\mathrm{d}y\ .
\end{equation}
The overall strength of these interactions is
controlled by $\alpha$ with their relative strength controlled by
$\gamma$. The interaction kernel takes the form
\begin{equation}
\hat{\kappa}(s)=\kappa_0-\kappa(s)\quad \text{for}\quad
\kappa(s)=\delta_{\beta}(s)
\end{equation}
where $\delta_{\beta}(s)$ is a general smooth function with the
properties
\begin{equation}
\delta_{\beta}(\pm\infty)=0\ ,\ \ \int\delta_{\beta} =1\ ,
\end{equation}
and the constant $\kappa_0$ is chosen so that $\int \hat{\kappa}=1$.
In this paper, we define $\kappa$ as
\begin{equation}\label{eq:kapbet}
\kappa(s)=\delta_{\beta}(s)=\frac{1}{2\beta}e^{-\frac{\vert
s\vert}{\beta}}\ ,
\end{equation}
although other choices could be considered. The new kernel $\hat{\kappa}(s)$ rewards proximity and thus represents an attractive water-water and water-head force, in
contrast to the original model of \citet{bap} which penalised
water-tail and head-tail proximity. The new parameter
$\beta$ controls the decay of the hydrophobic interaction, and will
be shown in \S\ref{subsub:epsibeta} to introduce a straightforward
analytical tool. (The notation was chosen so that the original
interaction kernel of \citet{bap} is a special case of ours, namely
$\kappa(s)=\kappa_0 - \hat{\kappa}(s)$ when $\beta=1$, but the
hydrophobic integrand is quite different.) 

The system is taken to be infinite with an averaged density $c_0$
for both $u$ and $v$. The original question addressed to the
paradigm in \citet{bap} was, in effect, whether the minimizer of the
energy functional $E$ favors aggregation or spreading when an excess
mass $m$ of lipids is added to the system, with all other parameters
fixed. Here, we concentrate on the utility of the BP paradigm, showing that physically realistic bilayer density profiles can be obtained from it numerically. We mainly focus on the results from our new model, for the reasons given in the introduction. To simplify the analysis and computation
we consider the case of periodic cells of length $2L$. All the
discussions are valid for the infinite system.

\subsection{Derivation of the Euler-Lagrange equations}\label{subsec:el}

Here we take the system to be incompressible, $p=\infty$, so that
\begin{equation}\label{eq:incomp}
1-u-v-\taum u -\taup v -w =0\ .
\end{equation}
The energy functional can be simplified as
\begin{equation}\label{eq:freesimp}
E_I=\int [\eta(u)+\eta(v)]+\alpha\int (1-u-v-\taum u-\taup
v)\hat{\kappa}*(1-u-v)
\end{equation}
subject to constraints
\begin{subequations}
\begin{align}
\label{eq:incompgeq} 1-u-v-\taum u-\taup v&=w\geqslant 0\ ,\\
\label{eq:masscons}\int (u+v-2c_0)&=m\ .
\end{align}
\end{subequations}
The result of scaling $T$ into $\alpha$ is that we can consider the
temperature effects by varying $\alpha$ (\S\ref{subsub:alpha}). We
have also dropped the entropy of the water molecules, which is
justified since although the entropy changes of the water associated
with reduced configurational arrangements around hydrophobic
moieties actually assists solvation, the effect is very small
\citep{kronbergetal1995}. Furthermore, we have taken $\gamma=1$,
effectively indicating that the electronegativities of waters and
heads are equal, and leaving the effects of the relative strength of
the water-water to water-head bonding preference to future work.

Using the definition (\ref{eq:kapbet}) of the interaction kernel and
the mass conservation constraint (\ref{eq:masscons}) the energy
functional becomes
\begin{equation}
\begin{split}
E_I=\int
[\eta(u)+\eta(v)]+\alpha\left(1-2c_0-\frac{m}{2L}\right)\int
(1-u-v-\taum u-\taup v)\\ -\alpha\int(1-u-v-\taum u-\taup
v)\kappa*(1-u-v)\ ,
\end{split}
\end{equation}
where we have chosen $\kappa_0=(2-e^{-L/\beta})/2L$. Using the
method of Lagrange multipliers we rewrite the energy functional as
\begin{equation}\label{eq:freewithlagrange}
E_T=E_I+\frac{K}{2}\int
\mu^2+\lambda_+\left[m-\int(u+v-2c_0)\right]+\lambda_-\left[\int
(u+v-2c_0)-m\right]\ ,
\end{equation}
where $K$ and $\lambda_\pm$ are Lagrange multipliers and
$\mu=(u+v+\taum u+\taup v-1)_+$, with $(\cdot )_+=\max\{\cdot,0\}$.

Carrying out calculus of variations in a formal way, assuming that
the order of integrations and translations can be changed wherever
necessary, we derive the Euler-Lagrange equations
\begin{subequations}\label{eq:el}
\begin{align}
\label{eq:elu} 0&=\log u-\alpha\kappa *(2u+2v+2\taum u+\taum v+\taup v)+K\mu +K\mu(x+\epsilon)+\lambda \ ,\\
\label{eq:elv} 0&=\log v-\alpha\kappa *(2u+2v+\taum u+\taup u+2\taup
v)+K\mu +K\mu(x-\epsilon)+\lambda\ ,
\end{align}
\end{subequations}
where
\begin{equation}\label{eq:lambda}
\lambda=\lambda_- -\lambda_+ +1+3\alpha-2\alpha(1-2c_0-m/2L)\ .
\end{equation}

We solve the Euler-Lagrange equations by first replacing them with
evolution equations based on gradient flows:
\begin{subequations}\label{eq:elgrad}
\begin{align}
\label{eq:elgradu} u_t&=-\log u+\alpha\kappa *(2u+2v+2\taum u+\taum v+\taup v)-K\mu -K\mu(x+\epsilon)-\lambda \ ,\\
\label{eq:elgradv} v_t&=-\log v+\alpha\kappa *(2u+2v+\taum u+\taup
u+2\taup v)-K\mu -K\mu(x-\epsilon)-\lambda\ ,
\end{align}
\end{subequations}
for the gradients $u_t=-(\delta E/\delta u),\ v_t=-(\delta E/\delta
v)$, with given values of $K, \lambda$. These equations are solved
numerically in the next section.

There are thus nine model parameters in total: seven apparently
physical and two ($K, \lambda$) strictly numerical. $K$ and
$\lambda$ will be chosen subject to stability, symmetry, and water
conservation considerations described shortly. Of the physical parameters we consider
herein $\epsilon,\beta,c_0,m,\alpha$, as described previously.

\subsection{Numerical scheme and solutions}\label{subsec:numerics}

Our numerical scheme solves the gradient flow equations
(\ref{eq:elgradu},b) to find the densities $u,v$ which minimize the
Euler-Lagrange equations (\ref{eq:elu},b). The approach uses a
discretized grid with finite difference formulae for $u_t,v_t$.
Both first- and second-order backward differencing was used, with
the results in close agreement. Generally, very small time steps
were required. The numerical domain was taken to be periodic and of
length $2L$, in contrast to \citet{bap} who used a finite domain
with small decay at the edges. In both approaches, $u,v$ can and do
deviate from $c_0$ at the boundaries. In tests using the original
model, our numerical scheme reproduced the profiles of \citet{bap},
as near as we can tell, given that their simulation conditions were
not fully specified. Typical runtime on a machine with two 2GHz AMD
Opteron 270 Dual Core procesors with 16GB of DDRS-667 SDRAM is four
minutes.

The values of the Lagrange multipliers $K,\lambda$ are not specified
by the model. Indeed, we require an explicit penalty term in the
algorithm, effectively replacing $\lambda_- -\lambda_+$ in $\lambda$
with $\lambda^*(\int(u+v-2c_0)-m)$. Within the range of values of
$K,\lambda^*$ for which the numerics are stable, $K$ is chosen large
enough to ensure $w\geqslant 0$ but no larger, and $\lambda^*$ is
chosen to ensure that the solutions are symmetrical, as expected.
Typically, $K$ is of the order of $10^3$ (but can be as large as
$\mathcal{O}(10^4)$), whereas $\lambda^*$ is around $10^2$.

\begin{figure}[t]
\begin{center}
\psfrag{t}{tail} \psfrag{h1}{head} \psfrag{h2}{head}
\psfrag{w1}{water} \psfrag{w2}{water} \psfrag{e}{$\epsilon$} \epsfig{file=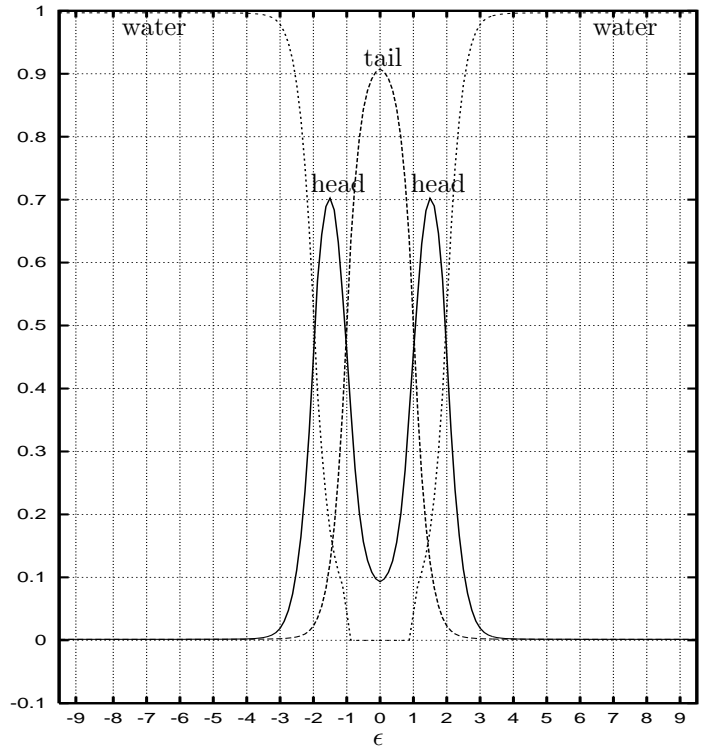,
height=10cm,width=10cm} \caption{A sample ``bilayer'' for the
parameter set $\alpha=3, \epsilon=2, \beta=1, c_0=0.024,
m=0.05*2L$.} \label{fig:samp02}
\end{center}
\end{figure}

A sample result is shown in figure \ref{fig:samp02}. The system has
separated into a well-defined bilayer-like profile, in which the
hydrophobic tail region is separated from the water region by two
peaks in the hydrophilic head group density. Although it is not our
purpose to compare solutions of our model with those of the original
model of \citet{bap}, we here note briefly that the forms are
similar, but with lipids being slightly more strongly drawn into the bilayer
of the new model. All the results presented in this paper are
independent of domain length $2L$ (above a certain value) and grid
resolution (beyond some level of coarseness).

\subsection{Smoothing and the choice of lipid model}\label{subsub:lipmod}
Referring to our discussion in \S\ref{sec:intro} of the formulation
of the BP paradigm, the densities of physical lipid moieties are smoothed spatially and temporally into the densities of the model ``lipids''. In general, there will not be a one-to-one correspondence between each such ``lipid'' and the physical lipid molecules, but there is a rigorous
correspondence between the densities of the one and those of the
other. We prefer to call the paradigm's ``lipid'' a \emph{model
lipid component}. These components represent an as-yet unclear
spatial smoothing (coarse-graining at the same length scale) and temporal smoothing of the
molecular-level information. Regardless, we are free to choose any
lipid component model; we have here picked the simplest. This does
not mean that any information from the physical lipid is lost, but
that it is smoothed into (some sum of the densities of) the two
beads per component which we have. This is important from the perspective of
integrating this paradigm into a multiscale simulation, for which we
are interested in general bilayer characteristics such as bending
rigidity which rely more on physical thickness rather than details
of the bilayer profile. A simple rod-and-two-beads model does the
job of capturing \emph{all} of the molecular data into a much
smaller number of variables, and in general no more complex model lipid
components need be considered. One exception might be to replace the
rigid rod by a spring, since especially in the presence of embedded
proteins physical lipid molecules can stretch considerably
\citep{lee2003}. However, even this case may be addressed by the
current model lipid component.

The temporal smoothing occurs
in moving from the discrete to the continuous description, working
at the lipid length scale. The continuum densities are identified
formally with the discrete ones over a sufficiently long time scale,
but it is not clear what this time scale is. This smoothing,
although unclear in detail, is central to the smoothed paradigm and
is important not just because it enables access to longer time scales and
hence can act as a bridge from a lower level simulation to a higher
level one, but also because it captures some detail of the real
thermal motion of the lipid molecules within the bilayer, which is a
defining characteristic of bilayers and plays a key role in their
function.

Together, the spatial and temporal smoothings are evident in the numerical results in
that the model lipid components have combined to create a bilayer of total width greater than the naively-expected $2\epsilon$, as can be seen in figure \ref{fig:samp02} (see a more formal discussion in \S\ref{subsec:newinterp}).

\section{Connecting the paradigm with the physics}\label{sec:phys}

\subsection{Physical bilayer properties}\label{subsec:bilayer}
Three biologically significant membrane characteristics are (1) the
elastic moduli, (2) the intrinsic monolayer curvature, and (3) the
bilayer thickness \citep{mouritsen2005}. The latter is the focus of
the present section ((1) and (2) require working in higher
dimensions).

The averaged thickness \dsz\ of the hydrophobic core, or saturation
zone, is a common physical measure of bilayer thickness. In a
physical system this can be increased by the following means
\citep[§8.3]{mouritsen2005}: increasing the length \lt of the tails;
replacing the double carbon bonds by single bonds in the tails; decreasing
the degree of hydration; increasing the cholesterol concentration;
decreasing the temperature. (The latter means each lipid has less
kinetic energy, making all lipids on average less convoluted, and
hence longer on average, while in a bulk effect lowering the
temperature makes the arrangement of lipids more crystalline. Both
effects increase \dsz.) These effects can increase \dsz\ by several
percent \citep[\S 9.2]{mouritsen2005}.

\begin{figure}[t]
\begin{center}
\epsfig{file=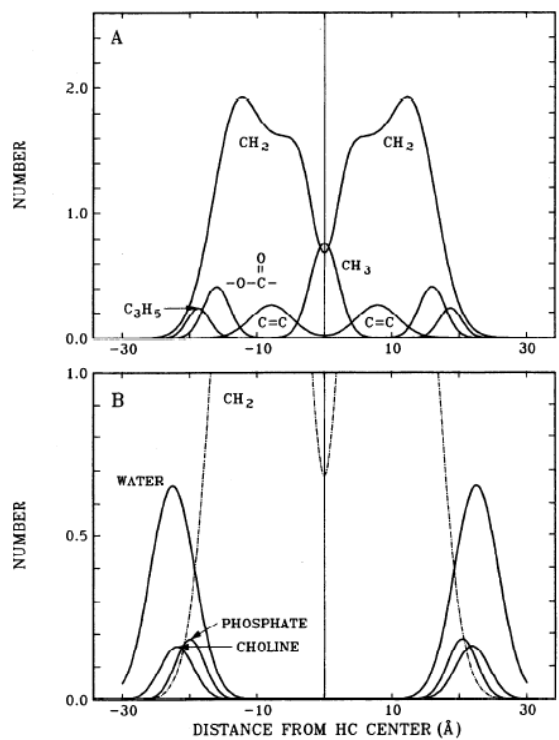, height=10cm,width=8.14cm}
 \caption{The experimentally-determined structure of a DOPC bilayer (figure
reproduced from \citet{wienerwhite1992} with kind permission of the authors and the Biophysical Society.)} \label{fig:pdf}
\end{center}
\end{figure}

The main approaches used in physical experiments to determine the
time-averaged structure of lipid bilayers, and hence their
thickness, exploit the high structural periodicity in the direction
normal to the bilayer, for example in combining diffraction data
from x-ray and neutron scattering \citep{wienerwhite1992}. One such
data set is represented in figure \ref{fig:pdf} for the DOPC lipid
molecule. In this figure, the water density measures only the waters
of hydration (those bonded to head groups).

\subsection{Making comparisons with physical meaurements}\label{subsec:newinterp}

For our analysis the key features of figure \ref{fig:pdf} are the
saturation zone, defined as that region where $w\approx 0$, the
density curves of the end of the tail (here, the $CH_3$ moiety of
DOPC), and those of the head group (here, choline and phosphate). The DOPC bilayer will form the basis of the remaining work
in this paper, but the key point is that any single-species
bilayer can be simulated by the BP paradigm, when basic
structural details are known from experiments.

We consider only those numerical results with a clear bilayer
structure like that shown in figure \ref{fig:samp02},
namely\label{page:clearbilayer} with a single saturation zone of
width \dsz, a single tail peak with exactly two transitions from
concavity to convexity, and two head peaks of equal width \dhz, each
likewise with exactly two transitions from concavity to convexity.
Macro-separation of the two species $u$ and $v$ can and does occur
without these criteria being met, but our analysis is here concerned
only with those profiles rigourously of this form. The head zone
width \dhz\ is defined in the following way. Using the left-hand head
peak, let $x_L, x_R$ be points immediately to the left and right of
the peak satisfying $h_x=0, h_{xx}>0$, \emph{i.e.} local minima,
with the local maxima of the head peak located at $x^*$. We then
find $x_1,x_2$ from $h(x_{1,2})=\frac{1}{2}(h(x_{L,R})+h(x^*))$ and
define $\dhz=x_2-x_1$. Defining the head zone width in this way as
running from the left-hand midpoint to the right-hand midpoint of
the peak is not unusual (\emph{e.g.} \citet[fig
8.1]{mouritsen2005}), but none of our conclusions is changed
significantly by defining it as $x_R-x_L$.

Lipid bilayers have different \dhz\ and \dsz\ depending on the lipids
of which they are composed. Specifically, the ratio \dhz/\dsz\
depends on the choice of lipid molecule (other factors such as
temperature being equal) and so characterizes the bilayer properties
for our purposes. For the DOPC bilayer of figure \ref{fig:pdf},
$\dhz/\dsz\approx 0.4$.

We now show that varying the key paradigm parameters of
\S\ref{subsec:modmodel} enables a solution to be found corresponding
closely to any desired physical bilayer, with the DOPC bilayer as
our example. The method of selecting the values of the parameters is
summarized in \S\ref{subsec:params}.

\subsubsection{$c_0$, $m$ and the critical micelle concentration}\label{subsub:cmc}

\citet{bap} called $c_0$ the ``background concentration'' and $m$
the ``excess lipids'' above the background concentration. This
terminology stems from the physical phenomenon known as the critical
micelle concentration (CMC), in which the monomer density of lipids
in solution only increases up to the CMC, beyond which the excess
lipids aggregate into ordered structures. The type of ordered
structure depends on the lipid geometry \citep{boal2002}. The BP
paradigm takes $c_0$ to be the background concentration to which an
excess quantity of lipids $m$ is added.

When implementing the paradigm numerically we face the problem that
the smoothing of \S\ref{subsub:lipmod} only reveals the lipid
species \emph{a posteriori} once the numerical profiles are compared
with experimental ones. Without knowing the species, and without
also knowing \emph{a priori} the temperature of the system, the CMC
cannot be known at the start of the simulation, and hence neither
can $c_0$ and $m$. Moreover, the CMC refers to an average lipid
density, which here would be
\begin{equation}\label{eq:barrho}
\bar{\rho}=\frac{2}{2L}\int u+v =\frac{m}{L} +4c_0\ .
\end{equation}
Consequently, the true background density is $m/L +4c_0^*$, where
$c_0^*$ is the value of $c_0$ for which aggregates first form (with
all other parameters fixed), and the total excess of lipids is
$8L(c_0-c_0^*)$. Since $m$ and $c_0$ combine to form the CMC, we can
numerically fix $m$ at a working value and vary $c_0$ only.

We note here that the above argument suggests that $u$ and $v$
should decay to $(m/L + 4c_0^*)/4$ far from the bilayer, but both
the periodic domain used in the present numerical simulations, and
the finite arbitrary value of $L$ used in the numerical method of
\citet{bap}, allow $u$ and $v$ to vary away from the background
concentration at the ``edges''.

\begin{figure}[t]
\begin{center}
\psfrag{d}{$d_{\text{HZ,SZ}}[\epsilon]$} \psfrag{r}{\dhz/\dsz$[1]$}
\psfrag{c}{$c_0 [1]$} \psfrag{32h}{$\dhz(3,2)$}
\psfrag{33h}{$\dhz(3,3)$} \psfrag{42h}{$\dhz(4,2)$}
\psfrag{32s}{$\dsz(3,2)$} \psfrag{33s}{$\dsz(3,3)$}
\psfrag{42s}{$\dsz(4,2)$} \psfrag{32}{$(3,2)$} \psfrag{42}{$(4,2)$}
\psfrag{33}{$(3,3)$} \epsfig{file=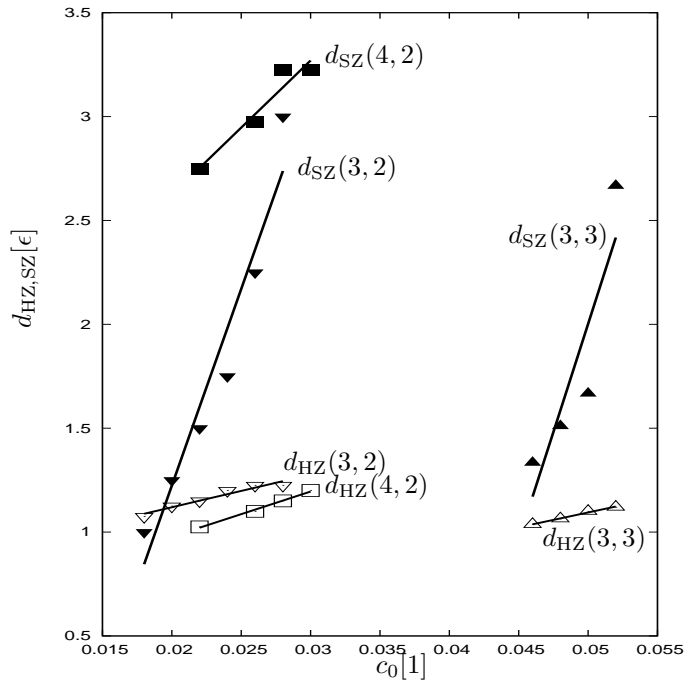,
height=9cm,width=9cm} \caption{\dhz\ and \dsz\
against $c_0$ for three different pairs $(\alpha,\epsilon)$, with
all other parameters equal. Straight lines of best fit have been drawn.} \label{fig:a2c_n_d}
\end{center}
\end{figure}

\begin{figure}[t]
\begin{center}
\psfrag{d}{$d_{\text{HZ,SZ}}[\epsilon]$} \psfrag{r}{\dhz/\dsz$[1]$}
\psfrag{c}{$c_0 [1]$} \psfrag{32h}{$\dhz(3,2)$}
\psfrag{33h}{$\dhz(3,3)$} \psfrag{42h}{$\dhz(4,2)$}
\psfrag{32s}{$\dsz(3,2)$} \psfrag{33s}{$\dsz(3,3)$}
\psfrag{42s}{$\dsz(4,2)$} \psfrag{32}{$(3,2)$} \psfrag{42}{$(4,2)$}
\psfrag{33}{$(3,3)$}\epsfig{file=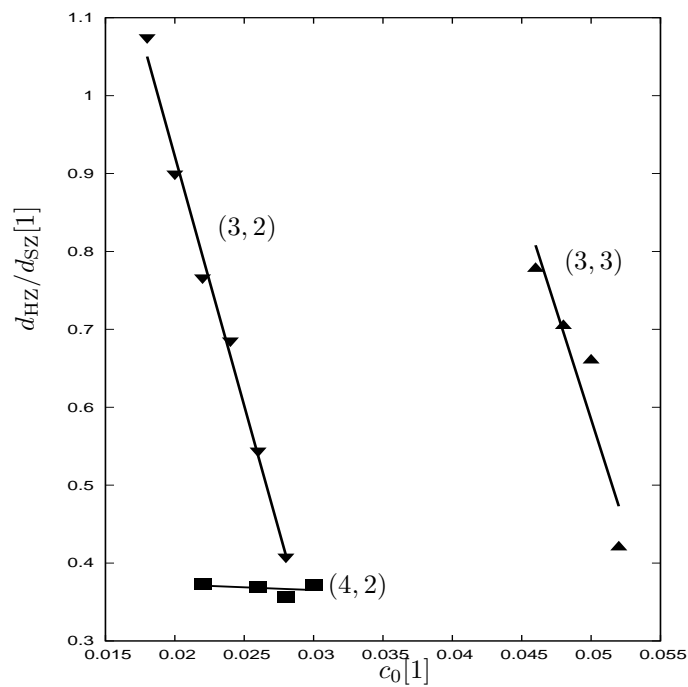,
height=9cm,width=9cm} \caption{The ratio
\dhz/\dsz\ against $c_0$ for the three parameter sets of figure \ref{fig:a2c_n_d}. Straight lines of best fit have been drawn.} \label{fig:a2c_n_d2}
\end{center}
\end{figure}

In figures \ref{fig:a2c_n_d} and \ref{fig:a2c_n_d2} we consider three data sets in which the
parameters $\alpha$,$\epsilon$, and $c_0$ vary, the values of
$\alpha$ and $\epsilon$ being given in parentheses on the graphs. The
widths \dhz\ and \dsz\ are normalized on $\epsilon$ in these figures.
Numerical data is represented by points and a straight line of best
fit is drawn in each case.

In the numerical system, as $c_0$ increases beyond $c_0^*$, the
excess lipids thus supplied should be drawn into the bilayer with
the numerical background concentration remaining more or less the
same. This required increase of \dhz\ and \dsz\ with increasing $c_0$
can be seen in figure \ref{fig:a2c_n_d}. Although this is
encouraging, this one-dimensional study cannot yet reveal the true
physicality of the models in terms of varying $c_0$. In a physical
system, as more lipids are added to the solution beyond the CMC,
more micelles (or other ordered structures) are formed, with the
average size of each barely affected \citep{boal2002}. Such a test
must wait for two-dimensional results.

More importantly, the data can be combined into the ratio \dhz/\dsz\
as shown in figure \ref{fig:a2c_n_d2}. Taking the
example of a DOPC bilayer, the data shows that we can choose
suitable parameter sets such that $\dhz/\dsz\approx 0.4$, namely
$(\alpha,\epsilon,c_0)=(3,2,0.028)$ for which $\dhz/\dsz=0.41$,
$(\alpha,\epsilon,c_0)=(4,2,0.022)$ for which $\dhz/\dsz=0.37$, and
$(\alpha,\epsilon,c_0)=(3,3,0.052)$ for which $\dhz/\dsz=0.42$. We
expect that ultimately $\alpha$ can be set physically (see also
\S\ref{subsub:alpha}) and that the role of $\epsilon$ is more one of
clarifying the structure (\S\ref{subsub:epsibeta}), so that
effectively here our only choice would be $c_0$, and the data shows
that a value can be chosen which yields a bilayer characteristic
close to the desired one. This is expected to hold for lipids other
than DOPC since there was nothing special about our choice of this
molecule.

\begin{figure}[t]
\begin{center}
\psfrag{d}{$d_{\text{HZ}}[\epsilon]$}
\psfrag{c}{$c_0 [1]$}
\epsfig{file=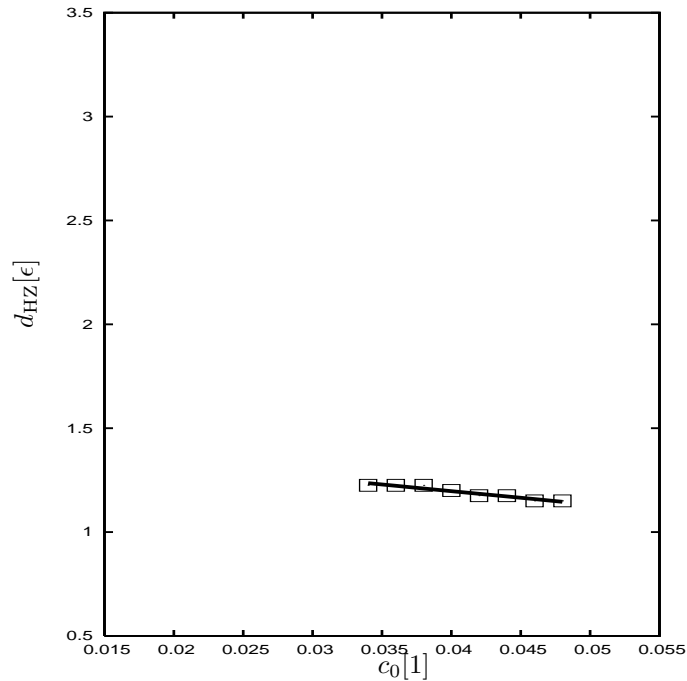,
height=9cm,width=9cm} \caption{Example data for the original model: \dhz\
against $c_0$ for $(\alpha,\epsilon)=(4,2)$, plotted on the same scale as figure \ref{fig:a2c_n_d}. A straight line of best fit has been drawn, indicating a non-physical inverse relationship between \dhz\ and $c_0$.} \label{fig:oldhz}
\end{center}
\end{figure}

Although it is not our intention to directly compare the new model with the original, it is worth noting here that the original model does not show the expected increase of \dhz\ with $c_0$: indeed, an inverse relationship between \dhz\ and $c_0$ holds for numerical data generated by the original model, as shown in figure \ref{fig:oldhz}.

\subsubsection{$\epsilon$ and the interaction decay length}\label{subsub:epsibeta}

That $\epsilon$ is not an actual lipid length has already been
discussed. What role, then, does it play?

Varying $\epsilon$ changes the degree of separation of the head and
tail regions, larger $\epsilon$ giving clearer bilayer structure.
With our definition (p.\! \pageref{page:clearbilayer}) of what
constitutes a realistic bilayer profile, we need only consider a few
order unity values of $\epsilon$; see \S\ref{subsec:params}.

Turning to the parameter $\beta$, which controls the interaction
decay length, we consider another special case of the kernel
function of equation (\ref{eq:kapbet}). Our ``short-range
interaction'' approximation takes $\beta \rightarrow 0$, so that the
kernel function approaches the delta function
\begin{equation}\label{eq:kapdelt}
\kappa(s)=
\begin{cases}
1\ &,\quad s=0\ ;\\
0 &,\quad s\neq 0\ .
\end{cases}
\end{equation}
The Euler-Lagrange equations (\ref{eq:el}) then become
\begin{subequations}
\begin{align}
0&=\log u-\alpha(2u+2v+2\taum u+\taum v+\taup v)+K\mu +K\mu(x+\epsilon)+\lambda \ ,\\
0&=\log v-\alpha(2u+2v+\taum u+\taup u+2\taup v)+K\mu
+K\mu(x-\epsilon)+\lambda\ ,
\end{align}
\end{subequations}
which in the case of $K\rightarrow\infty$ reduce to
\begin{subequations}
\begin{align}
\log u - \alpha (2u+2v+2\taum u+\taup v +\taup v)&=-\lambda\ ,\\
\log v - \alpha (2u+2v+\taum u+\taup u +2\taup v)&=-\lambda\ .
\end{align}
\end{subequations}
Separating $\lambda$ into a constant term plus a term dependent on
$c$ by rewriting (\ref{eq:lambda}) as
$\lambda=\bar{\lambda}+2\alpha(2c_0+m/2L)$, we look for constant
solutions $u=v=c$, obtaining
\begin{equation}\label{eq:lamcon}
\log c-4\alpha c =-\bar{\lambda}\ .
\end{equation}
Differentiating (\ref{eq:lamcon}) with respect to $c$ yields the
critical concentration $c_c$ as the first necessary condition for
the existence of a solution:
\begin{equation}\label{eq:ccrit}
c=c_c\equiv\frac{1}{4\alpha}\ .
\end{equation}
The short-range interaction tool thus gives a simpler way to find
the same result (\ref{eq:ccrit}) as \citet{bap}. To simplify our
discussion, we assume that $c_0=c_c=1/4\alpha$ and take
$\alpha\geqslant 1$.

\begin{figure}[t]
\begin{center}
\psfrag{lam}{$\bar{\lambda}$} \psfrag{c1}{$c_1$} \psfrag{c2}{$c_2$}
\psfrag{l}{$l/L$} \psfrag{e}{$E_D$} \epsfig{file=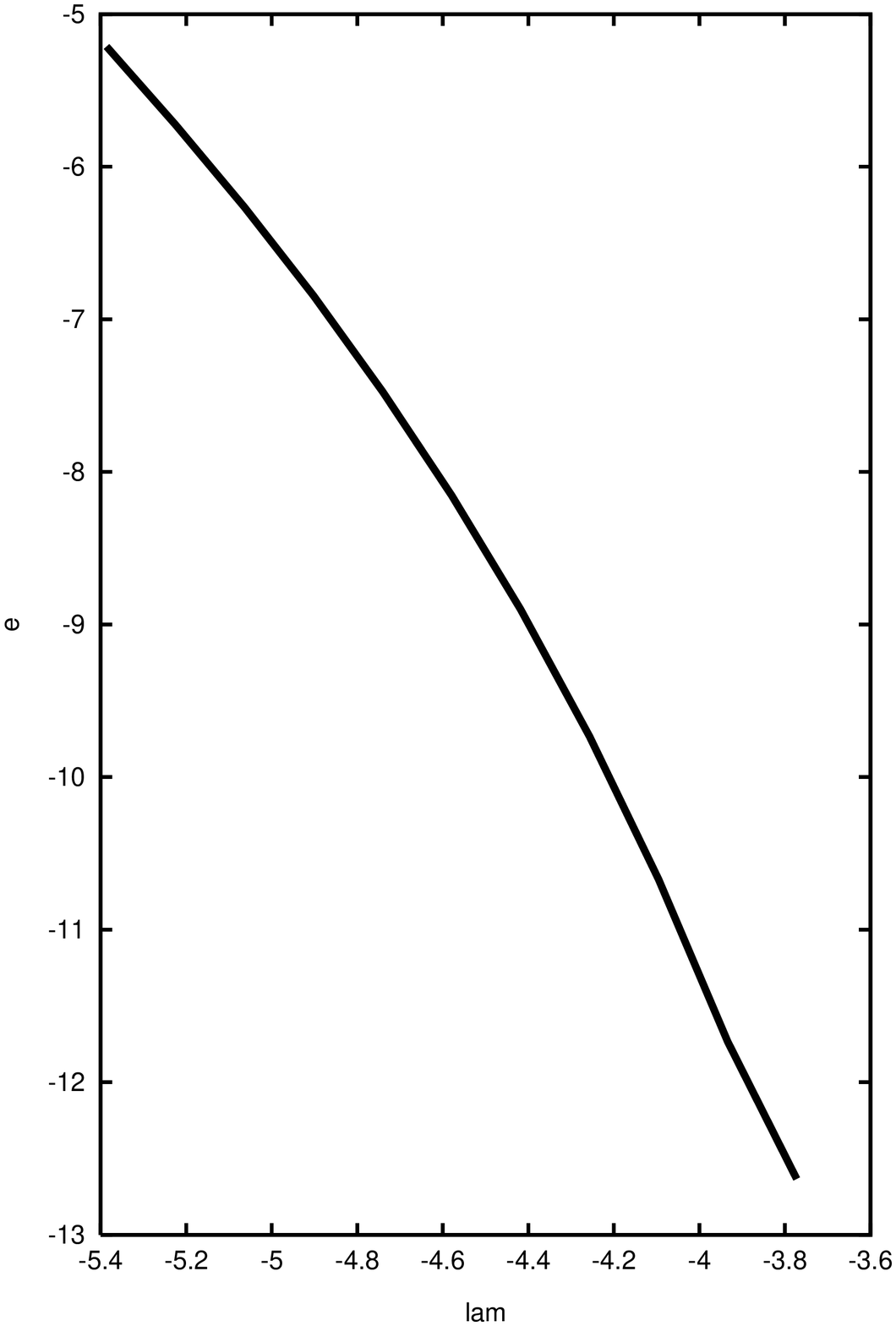,
height=6cm,width=6cm} \epsfig{file=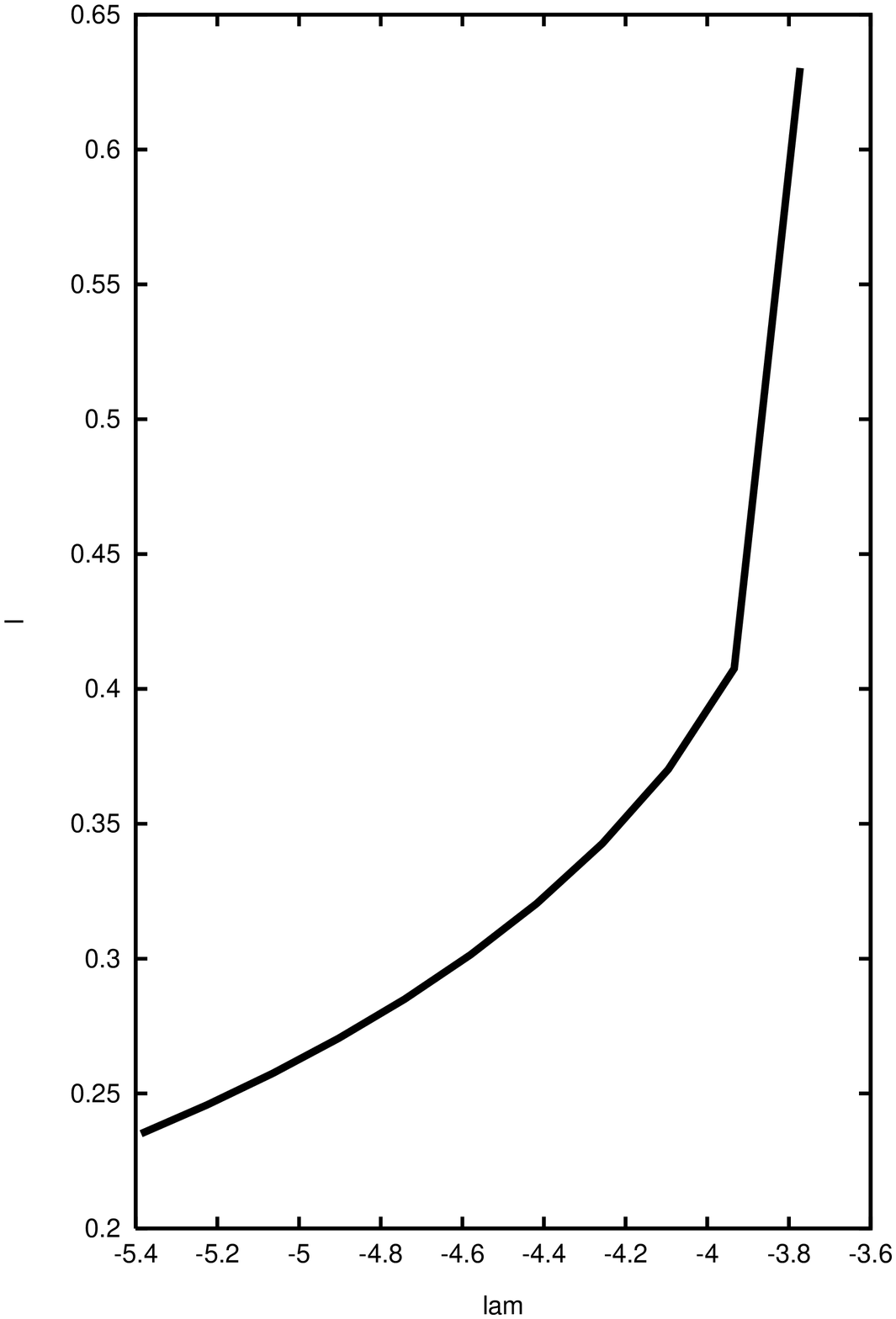, height=6cm,width=6cm}
\epsfig{file=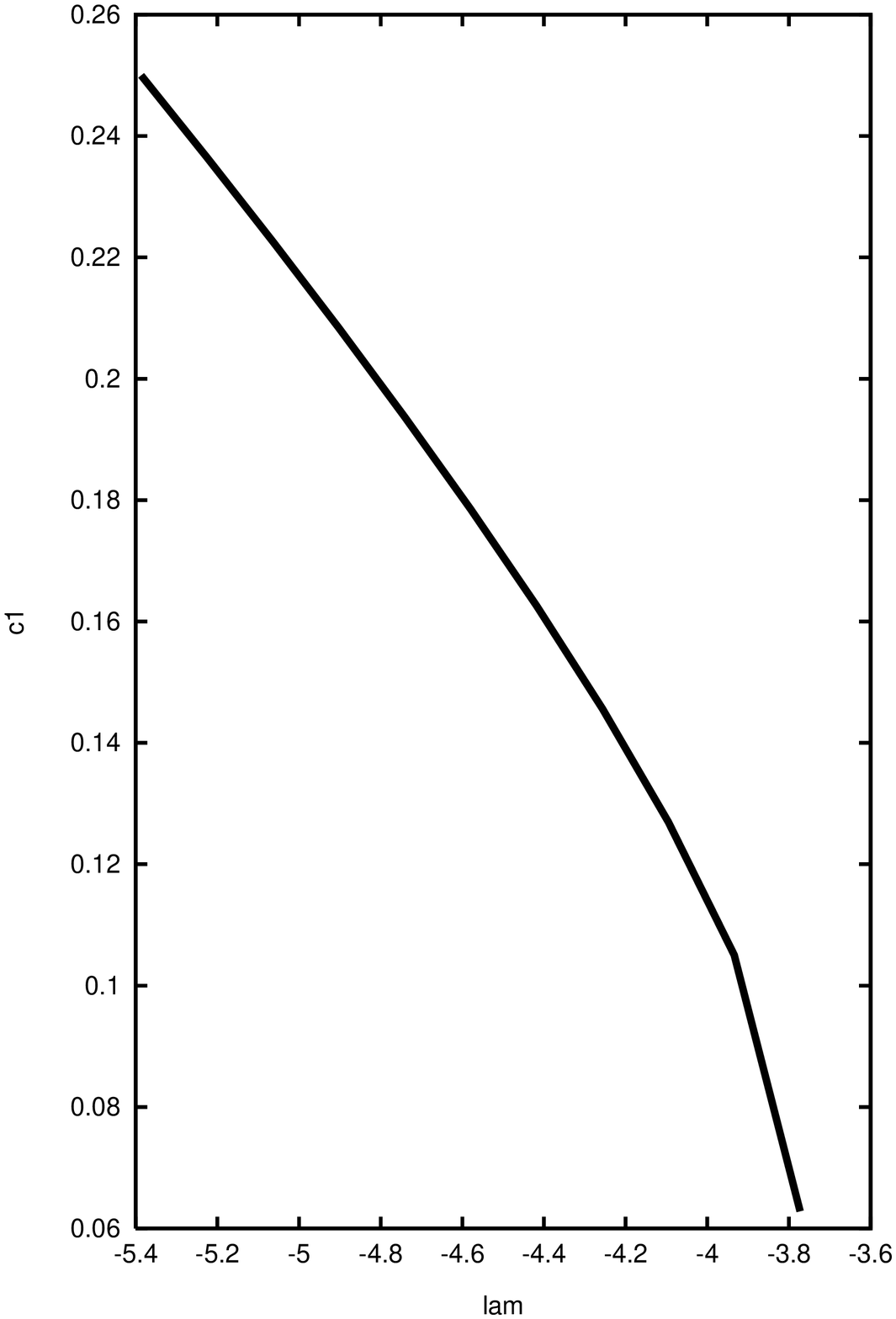, height=6cm,width=6cm}
\epsfig{file=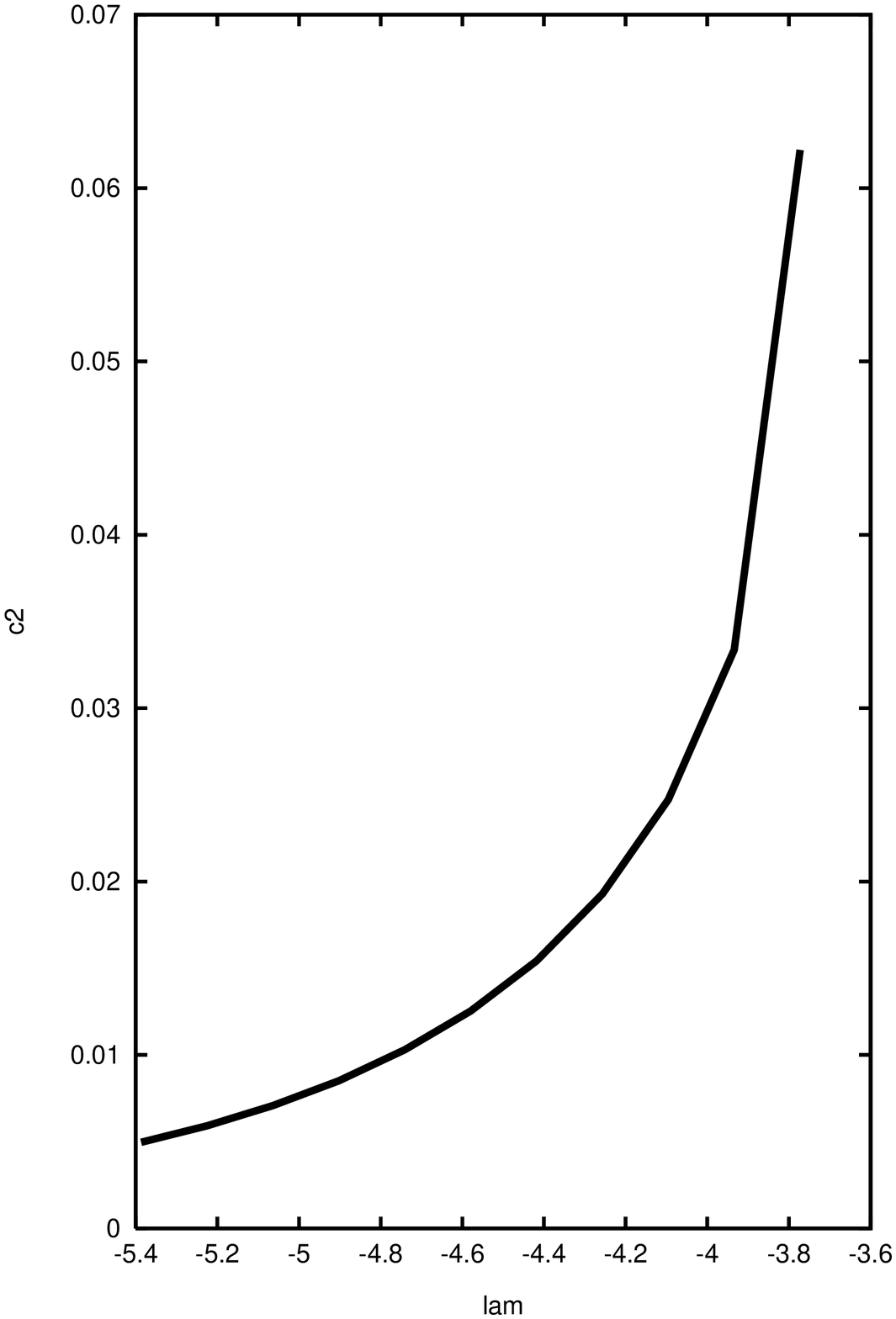, height=6cm,width=6cm} \caption{Solutions of
equations (\ref{eq:ed},\ref{eq:l},\ref{eq:lamcon}) for the
parameters $\alpha=4, \epsilon=2, c_0=1/(4\alpha), m=c_0/10, L=20$.}
\label{fig:bet0anal}
\end{center}
\end{figure}

The other necessary condition for the existence of a solution is
estimated as follows. From the incompressibility condition
(\ref{eq:incompgeq}) we see that $c\leqslant 0.5$. Substituting
$c=0.5$ into (\ref{eq:lamcon}) gives
\begin{equation}
\bar{\lambda}_{\text{min}}=-\log 2-2\alpha\ .
\end{equation}
When there is phase separation, (\ref{eq:lamcon}) must have more
than one solution: indeed, $\bar{\lambda}$ must also be less than
$\bar{\lambda}_{\text{max}}$ where
\begin{equation}
\bar{\lambda}_{\text{max}}=\log c_c-4\alpha c_c=-\log (4\alpha)-1\ ,
\end{equation}
in which case there is only one solution. When $\beta=0$ the
incompressibility condition implies $c\leqslant 0.25$, and therefore
\begin{equation}
\bar{\lambda}_{\text{min}}=-\log 4-\alpha\ .
\end{equation}
With
$\bar{\lambda}_{\text{min}}<\bar{\lambda}<\bar{\lambda}_{\text{max}}$
there are many possible states. The relevant question is whether
there exists a global minimum of the $\delta$-function free energy
\begin{equation}\label{eq:EDaslim}
E_D=\lim_{\beta\rightarrow 0} E_I=\int
\eta(u)+\eta(v)+\alpha\left(u+v-2c_0-\frac{m}{L}\right)(1-u-v-\taum
u-\taup v)\ .
\end{equation}
With $\bar{\lambda}$ chosen between its minimum and maximum values,
we assume that $u=v=c_1$ for $-l<x<l$ and $u=v=c_2$ elsewhere, where
$c_1$ and $c_2$ are two distinct solutions of (\ref{eq:lamcon}). We
also assume that the solution is periodic with period $2L$. The mass
conservation (\ref{eq:masscons}) gives
\begin{equation}\label{eq:l}
l=\frac{m+4L(c_0-c_2)}{4(c_1-c_2)}\ ,
\end{equation}
which in (\ref{eq:EDaslim}) yields
\begin{equation}\label{eq:ed}
\begin{split}
\frac{E_D}{4}=l\left[c_1\log c_1-c_2\log c_2+2\alpha\left(2(c_1+c_2)+2c_0+\frac{m}{L}\right)(c_1-c_2)\right]\\
+L\left[c_2\log
c_2+\alpha\left(c_0+\frac{m}{2L}\right)(4c_2-1)+4\alpha
c_2^2\right]-\epsilon\alpha(c_1-c_2)^2\ .
\end{split}
\end{equation}
In figure \ref{fig:bet0anal}, we have plotted $E_D$ as a function of
$\bar{\lambda}$, for $(\alpha,\epsilon)=(4,2)$. We can see that the
global minimum occurs when
$\bar{\lambda}=\bar{\lambda}_{\text{max}}$.
\begin{figure}[t]
\begin{center}
\psfrag{d}{} \psfrag{b}{$\beta$} \psfrag{h}{$\dhz [\epsilon]$}
\psfrag{s}{$\dsz [\epsilon]$} \psfrag{r}{$\frac{\dhz}{\dsz} [1]$}
\epsfig{file=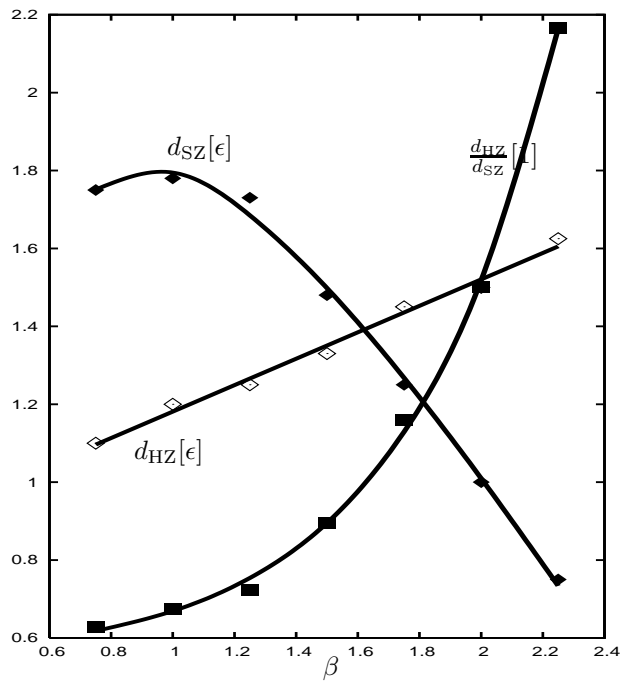, height=9cm, width=9cm} \caption{\dhz,
\dsz, \dhz/\dsz\ for different values of $\beta$. The other
parameters were fixed: in particular,
$(\alpha,\epsilon,c_0)=(3,2,0.024)$. For \dhz\ the drawn line of best fit
is straight $ax+c$, for \dsz\ it is of the form $ax+b/x+c$, and for
the ratio it is exponential and of the form
$a\exp(bx)+c$.}\label{fig:betvar}
\end{center}
\end{figure}

Returning to the question of finding a desired profile, figure
\ref{fig:betvar} shows how \dhz,\dsz\ and their ratio vary with
$\beta$ for all other parameters fixed. A multi-lammellar profile
began to appear for $\beta<0.75$, while separation without a
saturation zone appeared for $\beta>2.25$. Crucially, a bilayer-like
profile could be found for all values of $\beta$ in the given range,
meaning that the important ratio \dhz/\dsz\ can be fine-tuned to the
desired accuracy. See \S\ref{subsec:params}.

\subsubsection{$\alpha$ and the effects of temperature}\label{subsub:alpha}

Recalling the physical factors influencing \dsz\ listed in
\S\ref{subsec:bilayer}, since $\alpha$ is inversely proportional to
$T$, increasing $\alpha$ should increase the measurable \dsz.
Because this can be seen in figure \ref{fig:a2c_n_d}, it appears
that temperature-related mechanisms can be captured in 1D. Indeed,
turning to the incompressible free energy functional
(\ref{eq:freesimp}) in which $\alpha$ is scaled on $T$, increasing
$\alpha$ (decreasing $T$) reduces the effect of the entropy relative
to the interaction terms. This makes physical sense in that with
less kinetic energy the hydrogen-bond network is more strongly
preserved.

Further, the other physical temperature-related mechanism is that by
which lower temperatures straighten the lipid tails on average. (Although the
head groups also become ``straighter'' in the sense that they change
position less and therefore have a longer averaged shape, the main
effect is in the longer and more deformable tails.) As a result, we
would expect the ratio \dhz/\dsz\ of our model to decrease as
$\alpha$ increases ($T$ decreases), and this is seen in figure
\ref{fig:a2c_n_d2}.

Finally, we caution against one potential method of setting
$\alpha$. That increasing $\alpha$ (decreasing $T$) thickens the
physical bilayer (in terms of \dsz) is from the view point of moving
with the bilayer as it undergoes thermal fluctuations. But
increasing $\alpha$ (decreasing $T$) should suppress these thermal
fluctuations. If the model were capturing these undulations as well
then we might expect \dsz\ to decrease as $\alpha$ increased ($T$
decreased) because the order of magnitude of the thermal
fluctuations is larger than that of the changes in \dsz\ (a few
percent). Since we see the opposite, this precludes using the
amplitude of an averaged thermal undulation to fix $\alpha$.

\subsection{Choosing a parameter set}\label{subsec:params}

Based on \S\S\ref{subsub:cmc},\ref{subsub:epsibeta},\ref{subsub:alpha}, we can 
set the parameters to reproduce the profile of a desired lipid
species bilayer whose properties are known \emph{a priori}
from experimental data as follows.

Choose $\alpha=3$ or $4$, noting from figure \ref{fig:a2c_n_d2} that
from the numerical viewpoint $\alpha$ controls the sensitivity of
the ratio \dhz/\dsz\ to variation in $c_0$. Then, recalling that the
effects of $\epsilon$ and $\beta$ combine, we pick $\epsilon=2$ or
$3$, essentially only requiring clear separation of the head and
tail regions. Then we set $\beta=1$ to start. Next, since $c_0,m$
combine with $m$ fixed and $c_0$ to vary, we run the numerics with a
few $c_0$ until \dhz/\dsz\ is close to the desired value. These
results are then fine-tuned by varying $\beta$, which smoothly
varies the values \dhz\ and \dsz.

\section{Conclusions}\label{sec:conc}

We introduced a more physical model of the hydrophobic effect into
the continuum paradigm of \citet{bap} and showed that
one-dimensional numerical solutions can reproduce key
characteristics of physical lipid bilayers. The realised goal was to show that a more physical model enabling better connections with microscale numerics and first principles is capable within the paradigm of generating physically realistic bilayer density profiles. In particular, the mechanically-important bilayer thickness can be reproduced. Indeed, various key characteristics were shown to behave physically, at least one in contrast to the original model. The coarse-graining or
smoothing of the molecular information inherent in the paradigm was
clarified, and the smoothed numerical data calibrated to
measurements of physical bilayers.  Examining some of the key
parameters in turn, we gave a strategy for setting their values,
noting that future work, especially in higher dimensions, could make
this process even more robust by further appeal to physical arguments.  In
particular, $\alpha$ is formally linked to (the inverse of) the
temperature, and has been shown here to have that effect on the
numerical results, while $\beta$ should be related to the range of
the hydrogen bonding forces. Already, however, the new parameter
$\beta$ has been seen here to introduce the short-range interaction tool,
allowing analytical results to be obtained with greater ease.

Much work remains to show that the BP paradigm is a viable mesoscale
tool for multiscale simulations. The main aim is to work in higher
dimensions, and include compressibility effects by allowing $p$ to
vary. The former represents a significant computational challenge.
If the conclusions of the current paper are supported by
higher-dimensional work, then the paradigm can be made as physical
as desired by varying $\gamma$, considering other $\kappa$ (and
indeed whether ``promoting proximity'' is the best model of the
underlying physics), including electrostatic effects, weak head-head
van der Waals repulsion terms, and so on. The level of detail will
rest on computational issues, in particular cost-benefit
considerations in the light of the paradigm's use as a mesoscale
numerical filter.

\subsection*{Acknowledgements}
P.L.W. gratefully acknowledges the support of the Japan Society for the Promotion of Science and the University of Tokyo's Intelligent Modeling Laboratory during part of this work. His thanks for many helpful discussions go to members of the IKEMEN group of the Fluid Engineering Laboratory of the University of Tokyo. We also thank the RIKEN institute for their support through the project Research Program for Computational Science, R\&D Group for the Next-generation Integrated Simulation of Living Matters, which is supported by the Japanese Ministry of Education, Culture, Sports, Science and Technology.

\bibliographystyle{apalike}
\bibliography{wht}

\begin{thebibliography}{}

\bibitem[Ayton et~al., 2002]{aytonetal2002}
Ayton, G., Smondryev, A.~M., Bardenhagen, S.~G., McMurty, P., and Voth, G.~A.
  (2002).
\newblock Calculating the bulk modulus for a lipid bilayer with nonequilibrium
  molecular dynamics simulation.
\newblock {\em Biophys. J.}, 82:1226--1238.

\bibitem[Blom and Peletier, 2004]{bap}
Blom, J.~G. and Peletier, M.~A. (2004).
\newblock A continuum model of lipid bilayers.
\newblock {\em European J. Appl. Math.}, 15:487--508.

\bibitem[Boal, 2002]{boal2002}
Boal, D. (2002).
\newblock {\em Mechanics of the Cell}.
\newblock CUP, Cambridge.

\bibitem[Boryczko et~al., 2003]{boryczkoetal2003}
Boryczko, K., Dzwinel, W., and Yuen, D.~A. (2003).
\newblock Dynamical clustering of red blood cells in capillary vessels.
\newblock {\em J. Mol. Model.}, 9:16--33.

\bibitem[Chandler, 2005]{chandler2005}
Chandler, D. (2005).
\newblock Interfaces and the driving force of hydrophobic assembly.
\newblock {\em Nature}, 437:640--647.

\bibitem[Fraaije, 1993]{fraaije1993}
Fraaije, J. G. E.~M. (1993).
\newblock Dynamic density functional theory for microphase separation kinetics
  of block coploymer melts.
\newblock {\em J.\! Chem.\! Phys.}, 99:9202--9212.

\bibitem[Immergut, 1991]{immergut1991}
Immergut, E.~H., editor (1991).
\newblock {\em Encyclopedia of Applied Physics}, volume~18.
\newblock VCH, Berlin.

\bibitem[Kronberg et~al., 1995]{kronbergetal1995}
Kronberg, B., Costas, M., and Silveston, R. (1995).
\newblock Thermodynamics of the hydrophobic effect in surfactant solutions ---
  micellization and adsorption.
\newblock {\em Pure \& Appl.\! Chem.}, 67:897--902.

\bibitem[Lee, 2003]{lee2003}
Lee, A.~G. (2003).
\newblock Lipid-protein interactions in biological membranes: a structural
  perspective.
\newblock {\em Biochim. Biophys. Acta}, 1612:1--40.

\bibitem[Mchedlishvili and Maeda, 2001]{mchedlishvilimaeda2001}
Mchedlishvili, G. and Maeda, N. (2001).
\newblock Blood flow structure related to red cell flow: determination of blood
  fluidity in narrow microvessels.
\newblock {\em Jpn J. Physiol.}, 51:19--30.

\bibitem[Mouritsen, 2005]{mouritsen2005}
Mouritsen, O. (2005).
\newblock {\em Life - as a Matter of Fat}.
\newblock Springer, Berlin.

\bibitem[Sugii et~al., 2005]{sugiietal2005}
Sugii, T., Takagi, S., and Matsumoto, Y. (2005).
\newblock A molecular-dynamics study of lipid bilayers: effects of the
  hydrocarbon chain length on permeability.
\newblock {\em J. Chem. Phys.}, 123:184714.

\bibitem[Sugii et~al., 2007]{sugiietal2007}
Sugii, T., Takagi, S., and Matsumoto, Y. (2007).
\newblock A meso-scale analysis of lipid bilayers with the dissipative particle
  dynamics method: thermally fluctuating interfaces.
\newblock {\em Internat. J. Numer. Methods Fluids}, 54:831--840.

\bibitem[Vaidya et~al., 2007]{vaidyaetal2007}
Vaidya, N.~K., Huang, H., and Takagi, S. (2007).
\newblock Modelling {HA} protein-mediated interaction between an influenza
  virus and a healthy cell: pre-fusion membrane deformation.
\newblock {\em Math. Med. Biol.}, 24:251--270.

\bibitem[Wiener and White, 1992]{wienerwhite1992}
Wiener, M.~C. and White, S. (1992).
\newblock Structure of a fluid dioleoylphophatidylcholine bilayer determined by
  joint refinement of x-ray and neutron diffraction data.
\newblock {\em Biophys.\! J.\!}, 61:434--447.

\bibitem[Young, 2001]{young2001}
Young, D.~C. (2001).
\newblock {\em Computational Chemistry: A Practical Guide for Applying
  Techniques to Real-World Problems}.
\newblock John Wiley \& Sons, Inc., New York.

\end{thebibliography}

\end{document}